\documentclass[prl,pro,superscriptaddress,twocolumn,floatfix]{revtex4-1}
\usepackage{amsmath,amsfonts,amssymb,amsthm,bm,color,dsfont,graphicx,psfrag,mathtools,hyperref}%
\usepackage[inline]{enumitem}
\usepackage{appendix}
\usepackage[utf8]{inputenc}
\usepackage{csvsimple}
\usepackage{xcolor}
\usepackage{geometry}
\usepackage{mathrsfs}
\usepackage{longtable}
\usepackage{comment}
\newcommand{\etal}{\textit{et al}.}
\newcommand{\ie}{\textit{i}.\textit{e}.}
\newcommand{\eg}{\textit{e}.\textit{g}.}

\begin{document}

\title{Evidence for quasicritical brain dynamics}

\author{Leandro Fosque}
\email{lfosque@iu.edu}
\affiliation{Department of Physics, Indiana University, Bloomington, IN, 47405, USA}

\author{Rashid V. Williams-Garc\'{i}a}
\email{rwgarcia@iupui.edu}
\affiliation{Department of Mathematical Sciences, Indiana University-Purdue University Indianapolis, IN, 46202, USA}

\author{John M. Beggs}
\email{jmbeggs@indiana.edu}
\affiliation{Department of Physics, Indiana University, Bloomington, IN, 47405, USA}

\author{Gerardo Ortiz}
\email{ortizg@indiana.edu}
\affiliation{Department of Physics, Indiana University, Bloomington, IN, 47405, USA}

\date{\today}

\begin{abstract}
    Much evidence seems to suggest the cortex operates near a critical point, yet a single set of exponents defining its universality class has not been found. In fact, when critical exponents are estimated from data, they widely differ across species, individuals of the same species, and even over time, or depending on stimulus. Interestingly, these exponents still approximately hold to a dynamical scaling relation. Here we show that the theory of quasicriticality, an organizing principle for brain dynamics, can account for this paradoxical situation. As external stimuli drive the cortex, quasicriticality predicts a
    departure from criticality along a Widom line with exponents that decrease in absolute value, while still holding approximately to a dynamical scaling relation. We use simulations and experimental data to confirm these  predictions and describe new ones that could be tested soon.  
\end{abstract}

\maketitle


{\it Introduction.}
A cubic millimeter of cortex is showered with approximately $10^5$ synaptic inputs every second, even during sleep. The rate at which these synaptic currents arrive changes constantly, depending on cognitive state.  This situation drastically constrains any statistical analysis of living cortical networks. While equilibrium approaches might serve as approximations, nonequilibrium methods are required to accurately describe its dynamics \cite{Bonachela 2010}. Yet even the nonequilibrium framework of self-organized criticality \cite{Bak 1987, Jensen1998}, often invoked in these studies \cite{Eurich 2002,Levina 2009}, requires separation of timescales between a single incoming stimulus and the relaxation of the consequent avalanche. Such separation is not realistic in actual networks \cite{Priesemann 2014,WilliamsGarcia2014,WilliamsGarcia2017}. 

What is needed is an organizing principle for cortex dynamics that describes how these networks can appear to show some signatures of criticality despite receiving constantly changing external drive \cite{Priesemann 2018}. Work in \cite{Shew2015} recorded from turtle visual cortex while movies with changing images were delivered to the retina; work in \cite{Fontenele2019} recorded from freely moving and resting rats with changing levels of activity. In both studies, neuronal avalanches produced seemingly power law distributions with exponents that approximately followed a scaling relation \cite{Sethna 2001,Papanikolaou 2011}, indicating closeness to criticality. 

But what is perhaps most intriguing is that the exponents did not reveal a single universality class. Rather, the exponents could change across individuals, and even across time within an individual, all while approximately holding to a dynamical scaling relation. The data would seem to be at odds with the longstanding concept of a universality class \cite{ortiz}. 

Here, we address this apparent paradox with the principle of quasicriticality \cite{WilliamsGarcia2014}, which hypothesized that a healthy cortex always adapts to operate near a line of maximal dynamical susceptibility, and will be critical only in the limit of no external drive (stimulus, noise, or dissipation). It predicts that for nonzero drive, quantities like the dynamical susceptibility and information transmission
will no longer display singularities. As drive increases, the height of these peaks will decrease. Moving in parameter space along the (Widom) line of these maximal, but decreasing peaks, the avalanche (effective) exponents for size and duration as well as the branching ratio will decrease in testable ways (Fig. \ref{trend}). As the cortex moves along the Widom line of optimality, it also will remain close to the critical point, so adherence to a scaling relation is expected. Using these predictions, we will show that quasicriticality can account for the pattern of data seen in recent experiments \cite{Fontenele2019} without resorting to multiple universality classes. In addition, we show that our own experimental data fulfill quasicriticality’s specific predictions (Fig. \ref{trend}).

{\it Critical vs quasicritical vs non-critical. }
 It is important to highlight physical distinctions between the various hypotheses. 
\begin{figure}[h!] \hspace*{0cm}
    \includegraphics[width=0.7\columnwidth]{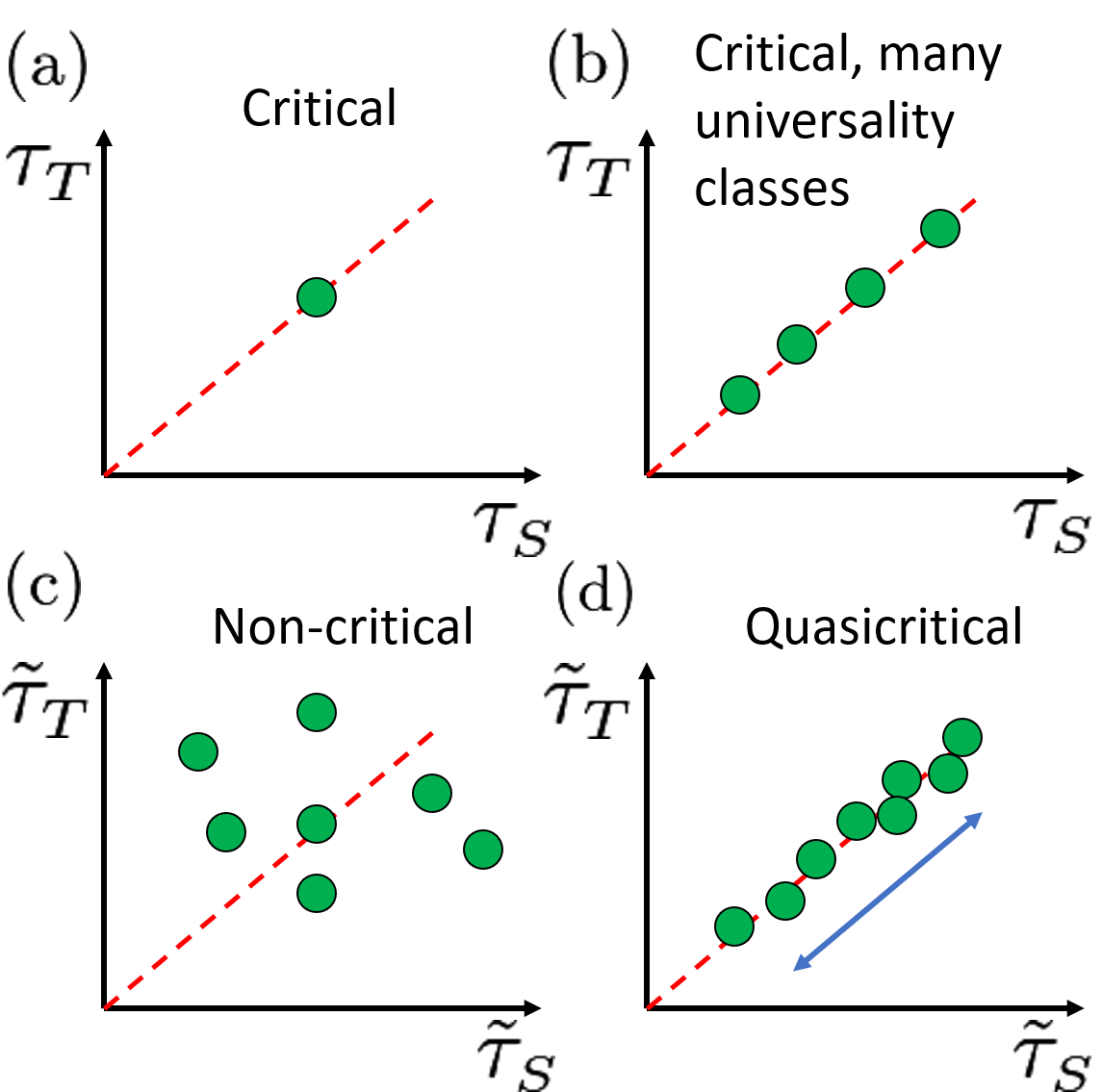}
	\caption{Exponent relations and dynamical scaling line, of slope $\gamma$,  for 
	the various physical hypotheses (see text).}
	\label{fig1}
\end{figure}
In neural activity data taken from cortical tissue, critical behavior can in principle be identified by examining dynamical scaling relations \cite{ortiz}. Power-law scaling permeates complex systems 
and a plethora of physical mechanisms generate scaling. In brain dynamics we  
argued \cite{WilliamsGarcia2014} that the relevant non-equilibrium scaling is associated to 
the well-known {\it absorbing-state phase transitions}, the most important class of which is 
directed percolation \cite{Henkel}. The finite-size scaling assumption for probability distributions of 
avalanches at criticality,
$P(q,L)=q^{-\tau_q}\Psi_q(q/L^{d_q})$ ,
establishes relations between the critical exponents $\tau_q$ and fractional dimension $d_q$, 
for quantities $q$ (e.g., avalanche 
size $S$ and duration $T$) given a system of linear size $L$, where $\Psi_q$ is the scaling function \cite{ortiz}. 
Assuming avalanches' size and duration satisfy scaling, then the conditional probability 
$P(S|T)\propto \delta(S-T^\gamma)$. This, in turn, leads to the conditional expectation value 
$\overline{\langle S \rangle}\propto T^\gamma$ with a resulting scaling relation  
$\gamma = (\tau_T-1)/(\tau_S-1)$, where exponents $\tau_T$ and $\tau_S$ are from 
avalanche duration and size distributions, respectively \cite{Henkel}. A pair 
$(\tau_S,\tau_T)$ defines a universality class \cite{Note} (see Fig. \ref{fig1}(a)). 
The criticality hypothesis of brain dynamics argues for such fine-tuning behavior. Moreover, from scaling one obtains bounds on critical exponents. For instance, the average avalanche size $\langle S \rangle \propto L^{\sigma_S}$ with 
$\sigma_S=d_S(2-\tau_S)$ if $\tau_S<2$, meaning that $\langle S \rangle$ diverges with system size, while $\sigma_S=0$ if $\tau_S>2$.

Figure \ref{fig1}(b) depicts multiple universality classes; there are several pairs of exponents, each dot fixed on the line \cite{Yaghoubi 2018}. If the data were not critical, pairs of exponents might be observed, but they would not generally adhere to the scaling relation (Fig. \ref{fig1}(c)). However, quasicriticality (Fig. \ref{fig1}(d)) predicts multiple pairs of exponents could be observed, all adhering to the scaling relation, and possibly moving along the line over time. Therefore, even though a system ideally belongs to a given universality class $(\tau_S,\tau_T)$\cite{Note}, 
the {\it measured effective} exponents $\tilde \tau_q$ may vary as a function of the stimulus or noise. 
A natural question is, how do these {\it effective} exponents change as a result 
and how far from the dynamical scaling line do they lie? This is a subject we explore next.

{\it Cortical Branching Model {\rm (CBM)}. } To address the question above we start analyzing the CBM \cite{WilliamsGarcia2014}. 
This is a non-equilibrium stochastic cellular automaton that captures main features of neural network data. Since we are interested in robust generic phenomena we need a flexible platform such as the CBM to answer that question. Certainly, 
the topology of the network dictates universality and the nature of the phases involved in its phase diagram. For our purposes, we focus on the case of strongly connected graphs, i.e., those with irreducible adjacency matrices \cite{SM}. In that case, 
we uncovered a rich phase diagram including a normal region with subcritical and supercritical phases separated by a critical (or crossover) region, and a quasiperiodic (chaotic) phase. 
Critical exponents were analytically determined using mean-field approximations and found to coincide 
with (mean-field) directed percolation universality class. Determining the universality class of the CBM in a strongly connected graph is an interesting question beyond the scope of this work \cite{note}. 
For more details we refer the reader to the original work \cite{WilliamsGarcia2014}.

We analyze the region in parameter space close to the Widom line, representing the crossover domain where the dynamical susceptibility associated to the density of active nodes (order parameter) attains its largest values. We consider random, strongly connected, networks of $L=128$ nodes, with a fixed in-degree of $k_{\sf in}=3$, branching parameter $\kappa$, with an exponential distribution of connection strengths \cite{SM}.

In the CBM the magnitude of the input activity to the network is represented by a probability of spontaneous activation $p_s$. The refractory period $\tau_r$, i.e.,  the number of time steps following activation during  which  a  node  cannot  be  made  to  activate, is a competing time scale in the problem.

As mentioned above, we want to explore the dynamical response in the domain surrounding the Widom line. To this end, for given values of $p_s$ and $\tau_r$, we locate the corresponding point along the Widom line by varying $\kappa$ until the dynamical susceptibility, at $\kappa_w$,  becomes maximal. It is precisely for this set of parameters that we next determine the effective exponents ($\tilde \tau_S, \tilde \tau_T$) for avalanche size and duration distributions. 
\begin{table}[h]
\begin{tabular}{|c|c|c|c|c|c|c|}
\hline
$\tau_r$ & $p_s$ &$\kappa_w$ & $\tilde \tau_S$        & $\tilde \tau_T$        & $\frac{\tilde \tau_T-1}{\tilde \tau_S-1}$     & $\gamma$ \\ \hline

1 & $10^{-3}$ & 1.10 & 1.57(1) & 1.82(1) & 1.44(3)            & 1.51(5) \\ \hline
1 & $10^{-4}$ & 1.12 & 1.64(2) & 1.99(4) & 1.55(8)             & 1.57(3)\\ \hline
1 & $10^{-5}$ & 1.17 & 1.69(2) & 2.14(15) & 1.65(22) & 1.66(1) \\ \hline\hline
5 & $10^{-3}$ & 1.19 & 1.62(2) & 1.85(8) &  1.37(14)            & 1.37(3) \\ \hline
5 & $10^{-4}$ & 1.23 & 1.67(1) & 1.99(12) & 1.48(18)             & 1.50(8) \\ \hline

\end{tabular}
\caption{Effective critical exponents for avalanche size and duration distributions as a function of $\tau_r$ and $p_s$.}
\label{table1}
\end{table}
Let us start by fixing $\tau_r=1$ and vary $p_s$ in the range $p_s \in [10^{-5}, 10^{-3}]$. The effective exponents, together with the computed values of $\kappa_w$,  are shown in Table \ref{table1}. As one can see, as $p_s$ increases, the exponents decrease in magnitude. Increasing the probability of spontaneous activity $p_s$, increases activity and, therefore, produces positive interference with avalanches, increasing their size and duration accordingly. At the same time, this induces a smaller magnitude of the effective exponents. 
Consider next increasing the refractory period to $\tau_r=5$. As appreciated in Table \ref{table1}, the optimal $\kappa_w$ is shifted to larger values. By increasing $\tau_r$, one reduces the number of possible nodes that could become active, which in turn reduces the size of the avalanches lowering the probability for the avalanche to continue spreading, and increases the magnitude of the effective exponents.
These constitute testable predictions of quasicriticality which could be checked, for instance, by application of drugs in living tissue resulting in an increase of the average refractory period.

Several comments are in order. First, although one could in principle establish a finite-size scaling extrapolation to 
$L \rightarrow \infty$ \cite{SM}, these effective exponents will never represent exact critical exponents since the system, when $p_s \ne 0$, is not scale invariant. Nonetheless, the effective exponents move along the dynamical scaling line. The universality class $(\tau_S,\tau_T)$, when $p_s\rightarrow 0$, is unique \cite{note} but depending on physical conditions the measured effective exponents wander along the dynamical scaling line. 
Second, the behavior just derived from computation corresponds to the optimal situation, that is, when the system operates exactly at the Widom line. On this line, one should obtain the best numerical agreement between the slope $\gamma$ derived from the effective exponents and the one obtained from the conditional expectation value $\overline{\langle S \rangle}$. 
Finally, as long as the system is close to the Widom line, that is, the quasicritical region \cite{WilliamsGarcia2014}, one will arrive at the same conclusions. But what happens if the system operates very far from the optimal region, such as in the deep subcritical region? \cite{SM}

{\it Signs of Quasicriticality in Experimental Data.} Before testing the quasicriticality hypothesis with empirical data, we need to make careful assumptions and apply strict criteria \cite{SM}. First, because living neural networks are open non-equilibrium systems, their state will fluctuate in two ways: (1) due to the inherent noise of external dynamic inputs and (2) the systematic fluctuations of the network response to those inputs. The result is fluctuations of the network state about a moving average in a multidimensional state space. This is realized, not only by dynamic external inputs, but also by the physiological responses of the network to those inputs, \eg, synaptic plasticity. In terms of the CBM, this means that 
the spontaneous activation probability $p_s$ and the branching parameter $\kappa$ will be dynamic---a situation not considered in our simulations. 

\begin{table}[h!]
\hspace*{-0.5cm}
\begin{tabular}{|c|c|c|c|c|c|c|}
\hline
Set & $\chi$ & $\sigma$ & $\tilde \tau_S$                     & $\tilde \tau_T$                     & $\frac{\tilde \tau_T-1}{\tilde \tau_S-1}$  & $\gamma$ \\ \hline
1  & 0.0096 & 0.7101   & 1.57(2)  & 1.73(4) & 1.29(3)  &  1.33(1)   \\ \hline
2  & 0.019  & 0.7322   & 1.58(3)  & 1.77(5) & 1.33(5)  &   1.33(2)  \\ \hline
3  & 0.0475 & 0.7433   & 1.65(6)  & 1.98(5) & 1.50(6)   &   1.48(7) \\ \hline
\end{tabular}
\caption{Effective exponents from empirical data. Sets (1) and (2) correspond to mouse 
cortical culture data and (3) corresponds to rat data.}
\label{table3}
\end{table}
According to our theory, however, fluctuations of the empirical values of $p_s$ and the branching ratio $\sigma$ (which we use as a proxy for the $\kappa$) induce fluctuations in the values of the effective critical exponents and hence the {\it apparent} universality class. Within a sufficiently short time frame, state fluctuations will be minimal; it is under these conditions that the network state can best be characterized as sub-, quasi-, or super-critical. 
Second, in order to satisfy quasicritical scaling, empirical data must be taken from or near the peak of maximum susceptibility, \ie, the quasicritical state \cite{WilliamsGarcia2014}.  

\begin{figure}[h!]
\centering
    \includegraphics[width=1\linewidth]{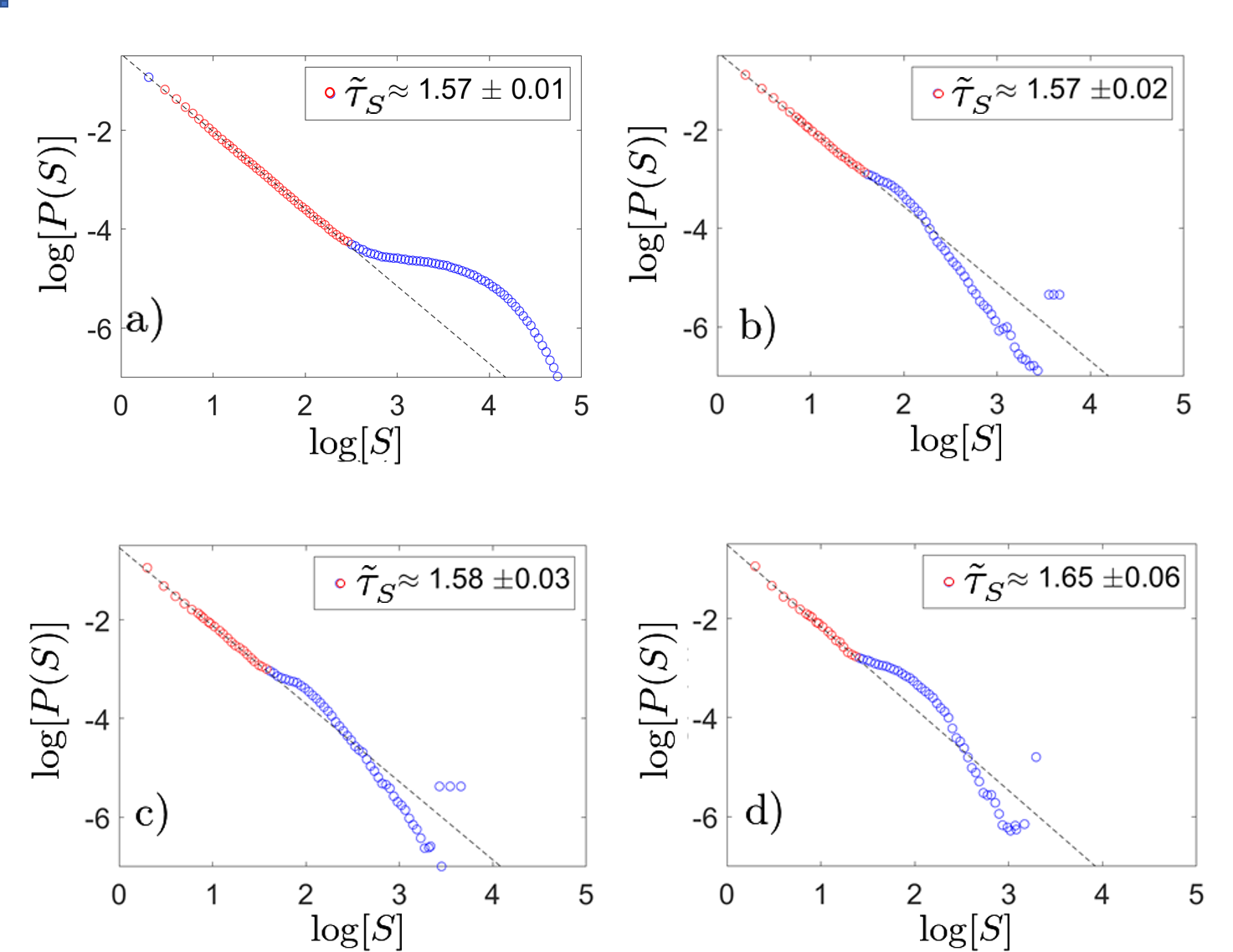}
    \includegraphics[width=1\linewidth]{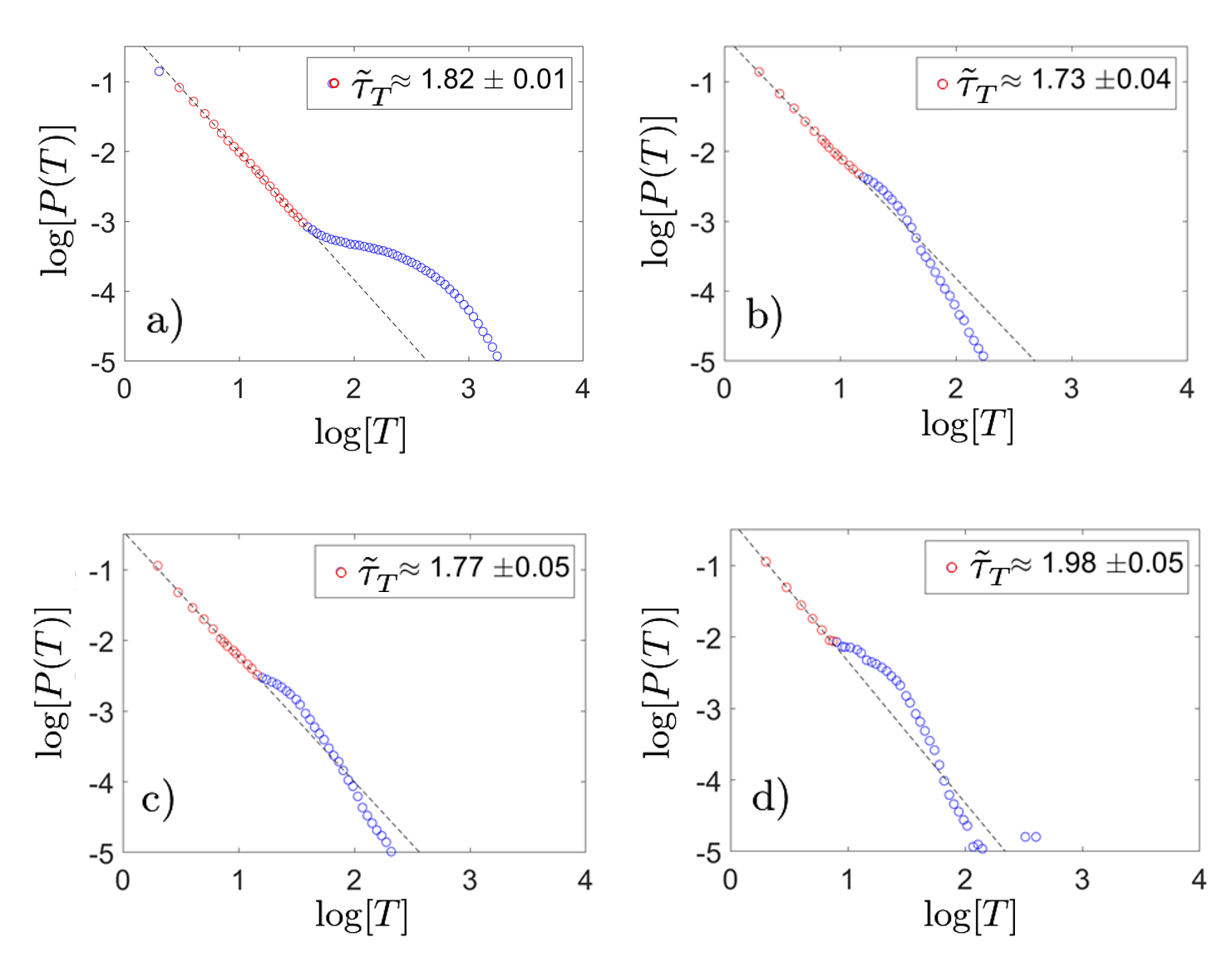}
    \caption{Logarithmically binned avalanche size, and duration, probability distributions \cite{Christensen2005} for (a) CBM simulation with $p_s=10^{-3}$, (b) Mouse data 23, (c) Mouse data 18, (d) rat data. Red circles were selected for fitting.}
    \label{distSize}
\end{figure}

A reliable method is then needed to quantify neural network state fluctuations, bin the data to adjust the time frame, and finally characterize the data. We start by quantifying these fluctuations in the language of statistical physics, introducing the {\it local time fluctuation} (LTF) of the network state. With a neural network of size $L$, we write the density of active nodes at time $t$ as $\rho_1(t)=1/L\sum_{i=1}^{L}\delta_{z_i(t),1}$, where $z_i\in\{0,1\}$ represents the state of quiescence or spiking, respectively, for neuron $i$. The mean population firing rate over a short time frame of $\Delta T$ time steps is then $\langle\rho_1(t)\rangle_t=1/{\Delta T}\sum_{t=1}^{\Delta T}\rho_1(t)$. The variability of $\rho_1(t)$ is the dynamical susceptibility \cite{WilliamsGarcia2014}, $\chi = L [\langle\rho_1^2(t)\rangle_t-\langle\rho_1(t)\rangle_t^2]$. Because $\chi$ varies dramatically depending on $L$, we normalize by the number of neurons to define the ${\rm LTF}=\frac{1}{\langle\rho_1(t)\rangle_t}\sqrt\frac{\chi}{L}$.
Effective exponents are then calculated from bins with similar, intermediate LTF values \cite{SM}. Alternatively, calculating  $\sigma$ and $\chi$ for each bin and mapping to LTF values shows that its intermediate values correspond to maxima of $\chi$ and intermediate values of $\sigma$. This is consistent with predictions made by our theory.

We analyze the dense microelectrode array recordings from rodent cortical tissue that were publicly posted on the CRCNS website \cite{CRCNS2016,CRCNS2016NMT}. For selection of specific data sets, we required a bump in the avalanche distributions when the data was in the supercritical regime (larger bins) and no bump when in the subcritical regime (smaller bins; see Fig. \ref{distSize}).  Although many data sets were consistent with quasicriticality, here we present only the 3 sets that satisfied these stringent criteria (for a clear description of all criteria involved in the analysis see \cite{SM}).  Table \ref{table3} lists results from the three data sets, along with values of the effective exponents. Data sets (1) and (2) are from mouse organotypic cortical cultures with number of neurons $L=310$ and $L=180$ respectively, binned at 1 ms. Data set (3) is from a rat organotypic cortical culture with $L=107$ and is binned at 5 ms. Although the rest of the data sets did not satisfy the stringent criteria, nonetheless, the determined effective exponents lie along the scaling line \cite{SM}. 

In agreement with our predictions, the values of $\chi$ and $\sigma$ from the data grow with the effective exponents. We report here branching ratios estimated from dividing the number of descendant neurons by the number of ancestor neurons. We found more advanced methods \cite{Wilting 2018} produced the same trends, but with smaller differences. Our experimental data (together with data from \cite{Fontenele2019}) and theoretical simulations are summarized in Fig. \ref{trend}.

\begin{figure*}[ht!]
    \centering
    \includegraphics[width=1\linewidth]{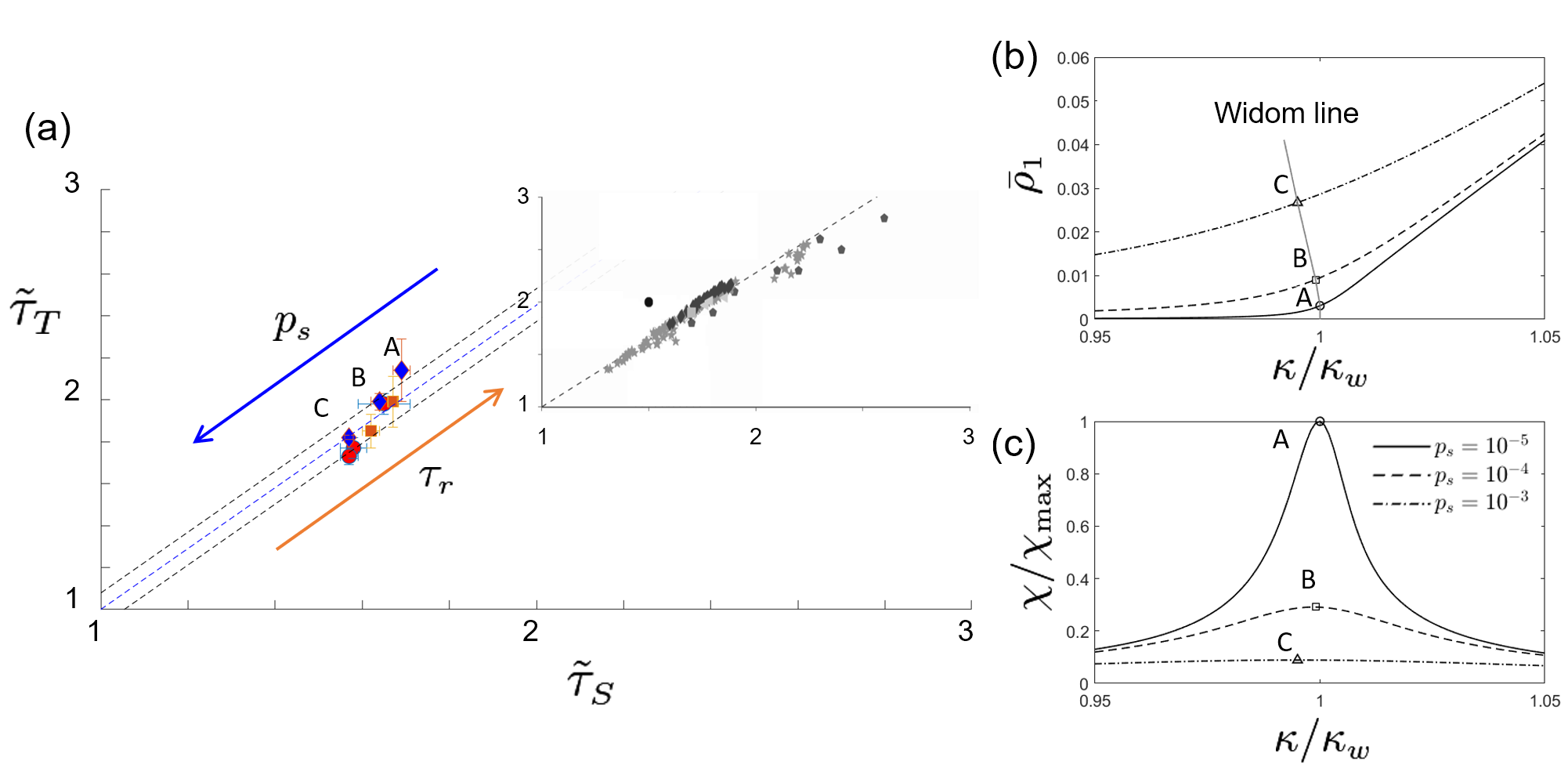}
    \caption{{\bf Predictions of quasicriticality}. (a) Simulations and data follow the scaling relation as predicted by quasicriticality. As $p_s$ is increased (blue arrow) exponents ($\tilde \tau_T$,$\tilde \tau_S$) decrease in the model (blue diamonds for $p_s=10^{-5},10^{-4},10^{-3}$) while still approximately holding to the scaling relation (dashed line). Note that susceptibility curves A, B and C in panel (c) correspond to these $p_s$ values.  Increasing $\tau_r$ (orange arrow) increases the magnitude of the effective exponents (the orange squares correspond to $\tau_r=5$ for $p_s=10^{-3}$ and $10^{-4}$ respectively). Actual data sets (red circles) also lie near the line, and have reduced values of susceptibility and branching ratio $\kappa_w$, in agreement with the simulations. Our scaling line has a slope of $1.45(4)$, in agreement with our data and differs from the slope in \cite{Fontenele2019} shown in the inset. Note that model parameters can be easily changed to match any slope, but that the trends in susceptibility and $\kappa_w$ as $p_s$ is increased will not change and constitute the specific predictions of the theory tested here. (b) Phase diagram showing how increased external drive leads to the Widom line; $y$-axis is density of active sites (order parameter), $x$-axis is branching ratio, $\kappa$ (control parameter). With minimal external drive ($p_s = 10^{-5}$), the solid black line at A comes near the critical point. As external drive is increased ($p_s = 10^{-4}$), the dashed line follows the solid black line, but rises above the critical point, with a gradual bend in the curve at B. Further increases in external drive ($p_s=10^{-3}$) cause the curve to rise more, bending near C. The thin gray line connecting A, B, C is the Widom line. Note that it tilts to the left, predicting that $\kappa_w$ should decrease as $p_s$ increases. (c) Susceptibility is maximal, but not divergent, along this line. Three plots of the susceptibility produced by simulations with different values of $p_s$. Note that increases in $p_s$ cause peak susceptibility to decline, leading to flattened curves.  }
    \label{trend}
\end{figure*}

{\it Outlook.} We have shown that quasicriticality can explain why cortical networks do not produce a single set of characteristic exponents as expected from a universality class. External inputs force these networks to operate in nonequilibrium conditions, away from a critical point. Yet the exponents produced under varying conditions still satisfy a scaling relation, indicating closeness to criticality. Quasicriticality predicts that increased drive will force cortical networks to depart from criticality, near the Widom line, preserving maximal susceptibility and information transmission. Moving along this line as drive is increased, the branching ratio at maximal susceptibility will decrease, and susceptibility will decrease. Simulations predict this, and the best data sets we have are consistent with these predictions. 

Future experiments must causally manipulate external drive, carefully tracking how the branching ratio and the susceptibility respond. These experiments have the potential to refute quasicriticality, or to further strengthen its standing. 


{\it Acknowledgment.} 
JMB gratefully acknowledges support from
NSF grant number 1513779 and Indiana University Bridge funds.


\begin{thebibliography}{99}
\bibitem{Note}
 Technically, in this case, one needs 
additional independent exponents to uniquely define the universality
class together with its corresponding scaling relation.
\end{thebibliography}
\end{document}